\documentclass[prb, twocolumn, superscriptaddress,showpacs,aps]{revtex4-1}

\usepackage{amsmath}
\usepackage{amstext}
\usepackage{amsfonts}
\usepackage{graphicx}
\usepackage{draftcopy}
\usepackage{flafter}
\usepackage[below]{placeins}
\usepackage{float}
\usepackage{tabularx}
\usepackage{times}

\def\ket#1{\left|#1\right\rangle}
\def\bra#1{\left\langle#1\right|}

\def\l{\ell}

\graphicspath{{figs/}}

\begin{document}

\title{Fast Quantum Control for Weakly Nonlinear Qubits: On Two-Quadrature Adiabatic Gates}

\author{Amrit De}
\affiliation{Department of Physics and Astronomy, University of California - Riverside, CA 92521}

\begin{abstract}
Adiabatic or slowly varying gate operations are typically required in order to remain within the qubit subspace in an anharmonic oscillator. However significant speed ups are possible by using the two quadrature derivative-removal-by-adiabatic-gate(DRAG) technique\cite{Motzoi2009,Gambetta2011}, where a second time derivative pulse component burns a spectral hole near an unwanted transition. It is shown here, that simultaneous optimization of the detuning and the pulse norm in addition, further reduces leakage errors and significantly improve gate fidelities. However, with this optimization accounting for the AC Stark shift, there is a low spectral weight pulse envelope regime, where DRAG is almost not needed
and where the two state error fidelities are stable against pulse jitter. Explicit time evolution calculations are carried out in the lab frame for truncated multi-level Transmon qubit models obtained from a tight-binding model.
\end{abstract}

\date{\today}

\maketitle

\section{Introduction}

Solid state qubits are typically formed from anharmonic systems such as defect centers\cite{Kane1998,Weber2010}, quantum dots(QD)\cite{Imamoglu1999,Petta2005,Hennessy2007} or Josephson junction (JJ) devices\cite{Devoret1989,Schoelkopf2008,Martinis2009} that have unevenly spaced energy levels. The two eigenstates that constitute the qubit, are selected by resonantly coupled to an orthogonal oscillating field which drives Rabi oscillations between the qubit levels.

 Since these qubits are not true spin-$1/2$ particles, driving the qubits too fast or non-adiabatically leads to leakages to other levels in these nonlinear systems,. The adiabatic theorem requires that in order to remain in an instantaneous eigenstate of the Hamiltonian, the drive should be slow or adiabatic. This however presents a huge problem as realistic qubits are not closed quantum systems and are subject to decoherence. Hence the qubit gate operations must be performed on a time scale that is much faster than the energy relaxation and the dephasing times, while also not leaking into unwanted levels. The error thresholds for the gate fidelities are about $10^{-4}$ to $10^{-6}$ for physical qubits\cite{Knill-error-bound,Steane-2003,Aliferis2006,Aharonov2006}.


The development of techniques for the efficient transfer of population to unpopulated atomic or molecular levels is a familiar problem in NMR and laser spectroscopy\cite{letokhov.book,Vitanov2001}. The optical Stimulated Raman Adiabatic Passage(STIRAP) scheme\cite{unanyan1998,sugny2007} is a well known for population transfer in a three-level $\Lambda$ system. Analogous schemes has been proposed for adiabatic passage through a triple quantum dot system using all all-electrical control\cite{Greentree2004,Greentree2004b}, manipulating neutral atoms in optical traps\cite{eckert2004} and for implementing qubit rotations\cite{Kis2002}. Typically in $\Lambda$ systems, the objective is to efficiently transfer the population from state $\ket{0}$ to $\ket{2}$ without populating $\ket{1}$.

In the context of quantum computing, in a ladder system, efficient population transfer is required between $\ket{0}$ and $\ket{1}$ without populating $\ket{2}$ which requires adibatic gating. Fast coherent qubit control has been proposed using resonant single-flux-quanta pulse trains\cite{Mcdermott2014}. Somewhat recently, it has been suggested that high fidelity qubit rotations with significant gate speed ups are possible by using the two quadrature derivative-removal-by-adiabatic-gate(DRAG) technique\cite{Motzoi2009,Gambetta2011,Motzoi2013}, where the second control pulse is proportional to the time derivative of the first. The reason why this works is because DRAG creates a spectral hole close to the unwanted transition frequency that causes leakage to the third level. This gating technique has been studied and adapted by others\cite{Poudel2010,Chow2010,Martinis2014}.

Consine pulses\cite{Martinis2014} and the more commonly used truncated Gaussian pulses\cite{Motzoi2009,Gambetta2011} have faster rise times when compared to an untruncated Gaussian pulses. And it is in these particular high spectral weight pulses where applying DRAG makes the most difference. There are additional known conditions to enhance fidelities such frequency chirping\cite{Motzoi2009,Gambetta2011} to account for the AC Stark shift. But this is difficult to implement. Analytic conditions to optimize fidelity are typically based on pertubative techniques such as the Magnus expansion or average Hamiltonian theory. Though these are accurate to some order in time they might not hold true till the end of the gating pulse.


One can ask if it gate fidelities at the threshold limit are attainable using simple optimization procedures and pulses that are just as fast, but much simpler to implement.
In this paper it is shown here that simultaneously optimizing the detuning (to account for the Stark shift) and the $\pi$ pulse's norm (to account for the shift in the momentum matrix element) leads to much faster gate times. This has been recently shown for cosine pulses\cite{Martinis2014}.  It is shown here that optimizing the detuning and the norm along with DRAG leads to further significant improvements in fidelity -- but this is the case only for Gaussian $\pi$ pulses with larger cut offs. If one considers cutoffs that make the envelope more adiabatic, then the inclusion of optimized DRAG only makes a very small difference because of lower overall spectral weights.


Another question that needs to be addressed is under what conditions is the commonly used \emph{three-level} model sufficient, since the energy gaps are similar in a weakly anharmonic oscillator. The DRAG technique can be applied to systems with more than one leakage transition\cite{Gambetta2011}. The resonance linewidth and the spectral weight distribution around it depends on the envelope. The addition of the derivative pulse envelope along the second quadrature, creates a spectral hole but raises spectral weights elsewhere. Any even(odd) parity state always directly couples to an odd(even) parity state in a symmetric potential. Moreover shorter pulses have broader linewidths which will more easily drive an initial state into a superposition of multiple states that are nearly resonant. The increased chances of these parasitic transitions requires a more detailed calculation of the population transfer fidelities with more levels.

All calculations here, are done for the Transmon qubit. However the gating procedures discussed in this paper are relevant for any qubit formed from a nonlinear ladder spectra. The qubit rotations just need to be implemented on a time scale that is much faster than the dephasing and relaxation times set by the decoherence mechanisms. Sources of decoherence for different types of qubits include hyperfine interactions\cite{Khaetskii2002,deSousa2003,Yao.prb.2006}, phonons\cite{Yu2002,Golovach2004,Stano2006}, cavity mode decay\cite{Pellizzari1995,Blais2004}, $1/f$ charge- and flux-noise noise\cite{Wellstood1987,Vandersypen2004,MacLean2007,Taubert2008,Clarke2008,Sendelbach2008,Paladino2014}.

Although superconducting qubits do not have the best coherence times, they are at the forefront for quantum computing because of their scalability and ease of control and fabrication. The Transmon is presently the SC qubit of choice as other SC qubits such as Flux qubits, Phase qubits and the Quantronium are susceptible to flux noise\cite{Yoshihara2006,Sank2012,Ithier2005} and the Cooper-pair-box(CPB) is susceptible to charge noise\cite{McDermott2009,Paladino2014}. However CPB variants like the Transmons and Xmons are mostly insensitive to charge noise due to their flat bands\cite{Koch2007pra}. The coherence times for Transmons have already approached the sub-ms range\cite{Weides2011,Rigetti2012}. It is not only important to gate the Transmon qubits as fast as possible but ensure that they are stable against pulse jitter.

In general, any practical quantum computer has to be fault tolerant, which requires that there be a way to either actively or passively detect and correct errors. Dynamical decoupling(DD)\cite{Kofman2001,Khodjasteh2005,Uhrig2007,Lee2008} and decoherence free subspaces\cite{Bacon2000,Brion2007,Lidar2008} are examples of passive error correction.
Whereas quantum error correcting codes(QECCs)\cite{Calderbank1996,Steane1996,shor-error-correct,Knill-Laflamme-1997,Kovalev2013pra} correct errors actively.
And there are additional merits to combining both DD and QECCs\cite{De2013PRL,De2014pra,De2015universal}. A number of recent experiments have demonstrated the working of QECCs\cite{Barends2014Nature,Kelly2015,Corcoles2015} using Transmon qubits. In order for QECCs to work the error rate per qubit should be finite. It has been recently suggested that leakage errors or two-state-errors are the most detrimental for QECCs\cite{Ghosh2013} such as the surface code\cite{Fowler2012prl,Fowler2012pra}.

Only two-state errors are considered in this paper while discussing gate fidelities.  A tight-binding model is derived for the Transmon which gives explicit eigenvalues and eigenfunctions -- which can be systematically truncated as required. The rotating wave approximation(RWA) is avoided here and the time-ordered unitary time evolution calculations are all done numerically in the lab frame. This is done for both the full tight binding model and the truncated models. As the detunings to account for the AC Stark shift get bigger as the pulses get shorter. One can argue that it is best to stay in the lab frame when dealing with large detunings.

The model and the method are discussed next, followed by a discussion of the two quadrature pulses, their power spectra and the main results from the unitary time evolution calculations.

\section{Model and Method}\label{sec.model}



Consider the Hamiltonian for an externally controllable Transmon qubit : $\mathcal{H} = \mathcal{H}_o+\mathcal{V}$, where
{\begin{eqnarray}
\mathcal{H}_o&=&\frac{\hat Q^2}{2C}-E_J\cos(\varphi)\\
\mathcal{V}&=&E_e\varphi
\label{H}
\end{eqnarray}}
where $\varphi=2\pi\Phi/\Phi_o$ is the superconducting phase difference across the junction, $\Phi$ is the magnetic flux through the loop, $\Phi_o=h/2e$ is the magnetic flux quantum, and $C$ is the junction-capacitance, $E_J=I_c\Phi_o/2\pi$ is the Josephson energy, $\hat{Q}$ is the charge operator. In the case of a current biased Josephson Junction, $E_e=I_e\hbar/2e$, where $I_e$ is the externally applied current.

Since $[\Phi,\hat{Q}]=i\hbar$ form a canonical pair and the charge operator $\hat{Q}=-i\hbar\partial/\partial\Phi$, in the long wavelength limit $\mathcal{H}$ can be mapped on to a tight-binding model using finite differences:
\begin{eqnarray}
{H}_o &=& -\tau(c_{k}^{\dagger}c_{k+1} +  c_{k}^{\dagger}c_{k-1}) + (2\tau-{u}_{k})c_{k}^{\dagger}c_{k}\\
{V}(t) &=& S(t)\Sigma_x\\
\Sigma_x &=& \varphi_{k}c_{k}^{\dagger}c_{k}
\label{eq:TB1Q}
\end{eqnarray}
where $c_k$ and $c_k^{\dagger}$ are the quasi-particle creation and annihilation operators for phase site-$k$, $\tau=E_C/a^2$, $E_C=(2e)^2/2C$ is the JJ's charging energy, $a=\Delta\varphi$ is the lattice constant, $u_k= E_J\cos(\varphi_k)$ and $S(t)$ is an externally applied time dependent signal. $\Sigma_x$ is the \emph{momentum matrix}.

\begin{figure}
\includegraphics[width=1\columnwidth]{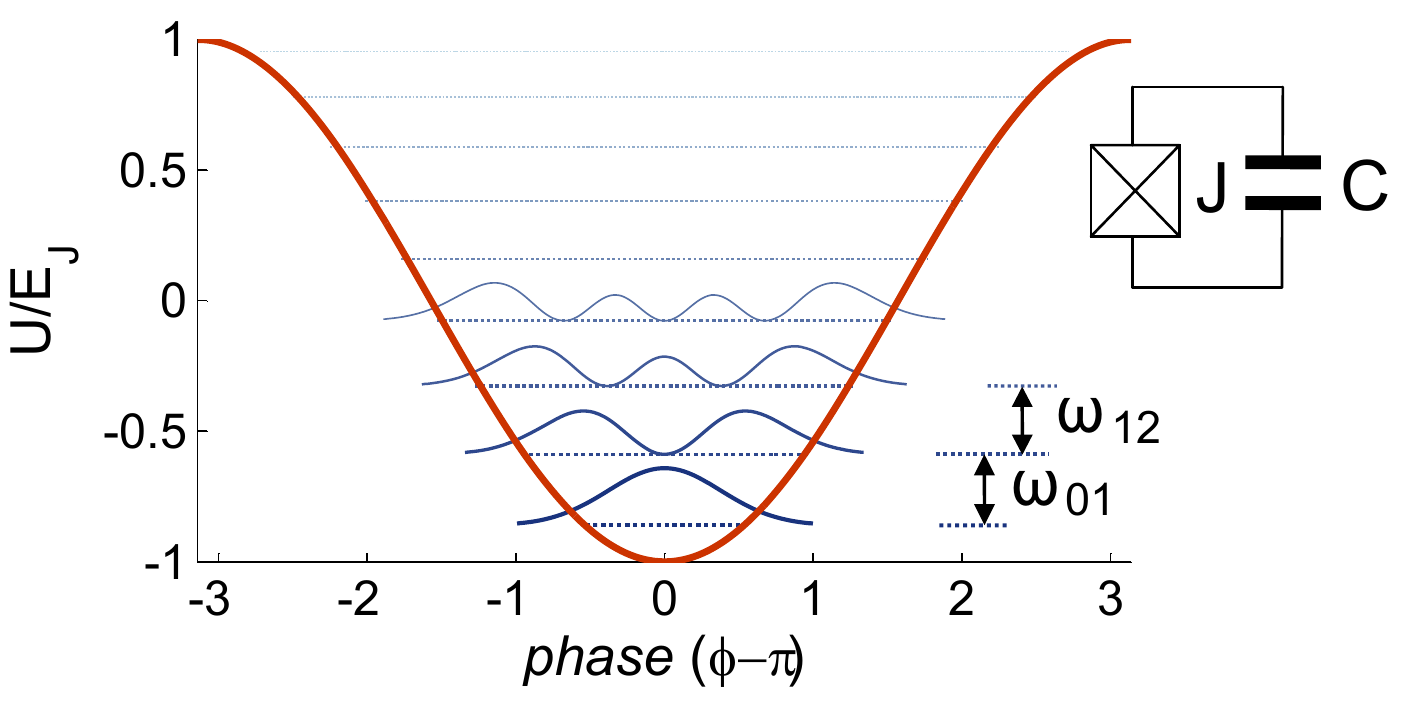}
\caption{Potential well, calculated energy levels and respective wave function amplitudes for the four lowest levels for the Transmon qubit. Here $E_J$ = 22.05 GHz and $E_J/E_C$=100 results in $\omega_{01}$ =6 GHz and $\Delta_2/\omega_{01}$=0.04. These calculations were done using 100 gridsites.}
\label{fig:well}
\end{figure}

The calculated energy levels and the lowest energy eigenstates used to from the qubit basis are shown in fig.\ref{fig:well}. $\omega_{01}=E_1-E_0$ and $\omega_{12}=E_2-E_1$, where $E_{0,1,2}$ are the lowest energy levels and $\Delta_2=\omega_{01}-\omega_{12}$. For the transmon qubit in order to simultaneously set $\omega_{01}$ =6 GHz and $\Delta_2/\omega_{01}$=0.04 we choose $E_J$ = 22.05 GHz and $E_J/E_C$=100.

For easier numerical calculations, one can construct a truncated model by projecting out the unwanted levels as follows:
\begin{eqnarray}
{H}_o' &=& \bra{\Psi'}H_o\ket{\Psi'}\\
V'(t) &=& D(t)\Sigma'_x
\label{eq:trunc}
\end{eqnarray}
\begin{eqnarray}
\Sigma'_x=\bra{\Psi'}\Sigma_x\ket{\Psi'}=\lambda_{ij}(\ket{i}\bra{j} + \ket{j}\bra{i})
\label{eq:sx}
\end{eqnarray}
where $i\neq j$ and $j>i$. If only three levels are considered, then $|{\Psi'}\rangle=\{\ket{\psi_0};\ket{\psi_1};\ket{\psi_2}\}$ and $\Sigma'_x=\lambda_j(\ket{j}\bra{j+1} + \ket{j+1}\bra{j})$. Here $\ket{\psi_j}$s are column vectors corresponding to the $j^{th}$ state of the full tight binding model. $\Sigma'_x$ is a symmetric pseudo-pauli $x$ matrix or a truncated momentum matrix. In a symmetric well any even(odd) parity state can directly couple to an odd(even) parity state. So for the three-level Transmon model, the ratio of the non-zero matrix elements are $\lambda_{12}/\lambda_{01}=\sqrt{2.08}$. If the fourth level is included, then $\lambda_{23}/\lambda_{01}=1.8$ and $\lambda_{03}/\lambda_{01}=0.008$.

For the Transmon or for any similar nonlinear multilevel qubit, one can construct pseudo-Pauli matrices from $\Sigma'_+=\lambda_{ij}\ket{i}\bra{j}$ and $\Sigma'_-=\lambda_{ij}\ket{j}\bra{i}$ where $j>i$, therefore
\begin{eqnarray}
\Sigma'_y&=&-i(\Sigma'_+-\Sigma'_-)\\
\Sigma'_z&=&[\Sigma'_x,\Sigma'_y]/2i
\label{eq:sz}
\end{eqnarray}
The two eigenstates that constitute the qubit, can be selected by applying an oscillating field along $\Sigma_{x(y)}$ which drives Rabi oscillations(RO) between the qubit levels and the RO frequency depends on the norm of the drive. $\Sigma_z$ can be useful for calculating chemical shifts and modeling interacting multi-level systems.

For the temporal dynamics, the wavefunctions need to be time evolved unitarily $\ket{\Psi(t')}=U(t',t)\ket{\Psi(t)}$, where the time evolution operator is
\begin{eqnarray}
U(t',t)= \hat{T}\displaystyle\int_{t}^{t'} \exp[-\frac{i}{\hbar}H(\tau)]d\tau.
\label{psipsi2}
\end{eqnarray}
$\hat{T}$ is the time ordering operator. The time evolution calculations here are carried out using a custom built C++ program based on the fourth-order Runge-Kutta algorithm for explicitly integrating the time dependent Schrodinger equation for unitary time evolution. Sub-picosecond time steps,$dt$, were chosen so that $dt\ll1/\omega$ where $\omega$ is the largest frequency component in the calculations. The resulting integration errors were better than or comparable to numerical precision. Note that all calculations are done explicitly in the {\it lab-frame}. If the qubit is initially in the state $\psi_\l(t)$, then the transition probability to a state $n$ at some later time is:
\begin{eqnarray}
\mathcal{P}_{\l\rightarrow n} = \left|\bra{\psi_n(t')}U(t^\prime,t)\ket{\psi_\l(t)}\right|^2
\label{eq:P0}
\end{eqnarray}
Rabi oscillations obtained from time evolving the {\it full} tight-binding model, using 100 gridsites and a slow drive, were in excellent agreement with that from a truncated 3-level model.


\section{Pulse Shapes and Power Spectra}\label{sec.spectra}

The state transfer probabilites can be also be understood in terms of the adiabatic theorem. Assume that at time $t$ the wavefucntion has evolved into $\psi_\l(t)$ which is different from $\psi_\l(0)$. The leakage probability is then approximately:
\begin{eqnarray}
\mathcal{P}_{\l\rightarrow n} &\approx& \frac{\hbar^2\left|\bra{\psi_n(t)}{\partial H}{/\partial t}\ket{\psi_\l(t)}\right|^2}{\omega^4_{\l n}(t)}
\label{eq:P1}
\end{eqnarray}
Hence the adiabatic theorem states that a physical system remains in its instantaneous eigenstate (at time $t$) only if the time dependent Hamiltonian changes slowly or external perturbation acting on it is slow enough -- which is the main design principle behind gating nonlinear systems. However more intuitive insights into the state transfer probabilities can also be obtained by examining the spectral weight distribution in the pulse's power spectra.

\begin{figure}
\centering
\includegraphics[width=0.9\columnwidth]{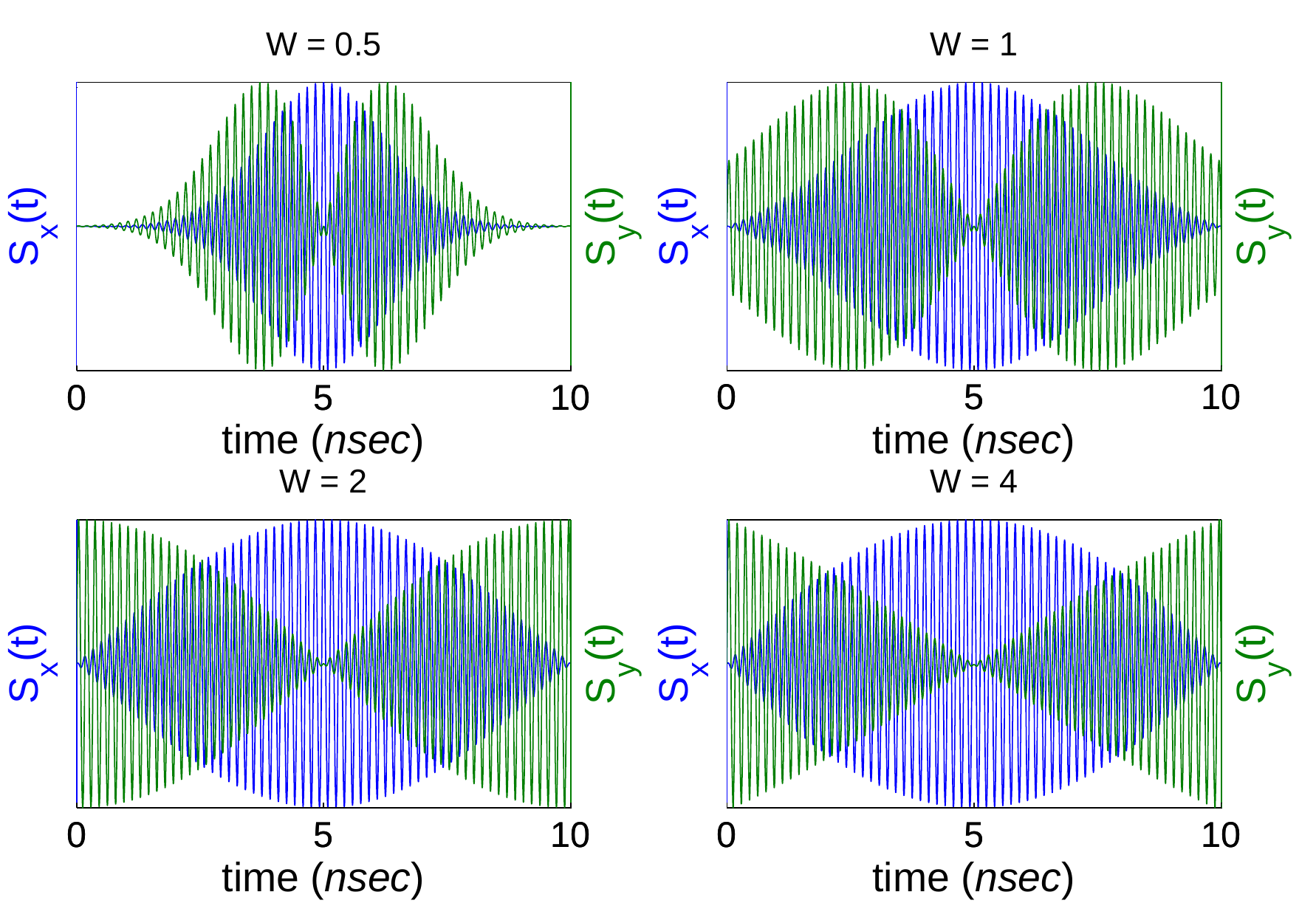}
\caption{ Gaussian pulse shapes and their derivatives in the lab frame for different pulse cutoffs $\exp(-2/W^2)$}
\label{fig:gauss}
\end{figure}

Consider a $\pi$ pulse that is resonant between levels 0 and 1 in the lab frame
\begin{eqnarray}
S_x(t) = A_x\frac{\pi}{\bra{0}\Sigma_x\ket{1}}\cos(\omega_{01}t)\xi(t)
\label{eq:Sx}
\end{eqnarray}
where $\xi(t)$ is the pulse envelope. For a Gaussian pulse of unit norm:
\begin{eqnarray}
\xi(t) = \frac{1}{\gamma\sqrt{\pi}} \frac{\left[{\exp\left( -[t-t_p]^2/\gamma^2 \right)} -\mathcal{G}\right]}{erf(\sqrt{2}/W)-t_p\mathcal{G} }.
\label{eq:G}
\end{eqnarray}
Here $t_p$ is the pulse width, $\gamma={W}t_p/2\sqrt{2}$ and $W$ is a parameter that determines the pulse cut-off, $\mathcal{G}=\exp(-2/W^2)$ and $erf(x)=\frac{2}{\pi}\int_0^x \exp(-t^2)dt$ is the error function. Whereas for cosine pulses and shaped pulses\cite{Pryadko2008PRA,Martinis2014} in general:
\begin{eqnarray}
\xi(t)=\frac{\theta}{t_p} + \frac{2\pi}{t_p}\displaystyle\sum_k\alpha_k\cos(\frac{2k\pi t}{t_p}).
\label{eq:cos}
\end{eqnarray}
where $\theta$ is the pulse rotation angle and $\alpha_k$ are the fitting parameters. For the usual $1-\cos$, $\pi$ pulse: $\theta=\pi$, $\alpha_0=-1$ and $\alpha_1=1/2$.

For regular pulses, if the pulse is too short then they become diabatic and this leads to leakages out of the qubit subspace. DRAG partially fixes the problem by constructing a multi-dimensional pulse by adding the derivative along $y$ in the RWA. The full signal will now consist of
\begin{eqnarray}
S(t) &=& S_x(t) - S_y(t)
\label{eq:St}
\end{eqnarray}
where
\begin{eqnarray}
S_y(t)= A_y\frac{\pi}{\bra{0}\Sigma_x\ket{1}}\left(\frac{|\bra{1}\Sigma_x\ket{2}|^2}{4\Delta_{2}} \sin(\omega_{01}t)\frac{d\xi(t)}{dt}\right)~~~~
\label{eq:Syt}
\end{eqnarray}
here $A_{y}$ is an adjustable parameter.



\begin{figure}
\centering
\includegraphics[width=1\columnwidth]{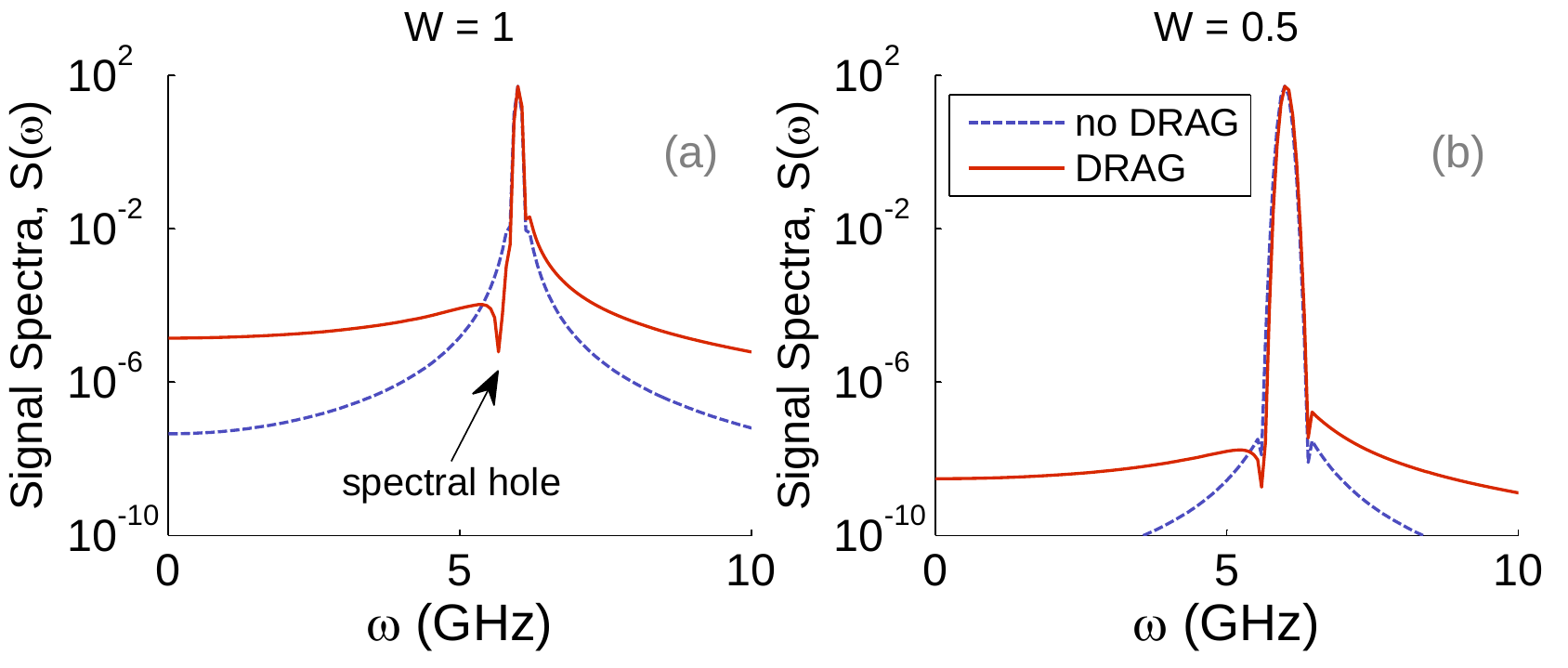}
\caption{ Fourier spectra of $t_p=15$ nsec pulse signals with DRAG ($A_y=1$) and without DRAG in the lab-frame with $\omega=6$ GHz  for cutoffs of (a) $W=1$ and (b) $W=0.5$. }
\label{fig:fft1}
\end{figure}

In fig.\ref{fig:gauss} Gaussian pulses and the derivatives are shown on a two axis plot for different pulse cutoffs (see Eq.\ref{eq:G}). The corresponding Fourier transform of the pulses are shown in Fig.\ref{fig:fft1} for half-DRAG ($A_y=1$) and without DRAG for two different cutoffs of $W=1$ and $W=0.5$ for a $t_p=15$ nsec pulse. The drive is modulated at $\omega=6$ GHz which is apparent from the plots. In the case of $W=1$ the addition of half DRAG leads to the creation of spectral hole close to $\omega-\Delta_2$ (more like $\sim\omega-1.5\Delta_2$ for $4\%$ nonlinearity). Therefore when comparing the cases with- and without DRAG, one can expect an improvement in spin-flip fidelity that is proportionate to the net spectral dip.

Now when considering a $W=0.5$ pulse one sees that even though the resonance-linewidth above $10^{-2}$ is atleast twice as broad as that of the $W=1$ pulse, the over all spectral weight away from resonance is smaller by four orders of magnitude. More over DRAG does not lead to the creation of significant net dip for $W=0.5$. DRAG does however raise the overall spectral weight for both the $W$ pulses and distributes it somewhat uniformly within the frequency window. This is true even in the case of $W=0.5$ where $S_y(t)$ does not begin as abruptly(see fig.\ref{fig:gauss}). This obviously calls for a full quantum multi-level time evolution calculation in order to better understand its effects on the qubit subspace.

In Fig.\ref{fig:fft1} the power spectra of $A_y=1$ DRAG pulses is shown for pulsewidths of $t_p=10$ nsec and $t_p=30$ nsec for various $W$ cutoffs. As longer pulses and smaller $W$ pulses are more adiabatic, they have much smaller spectral weights as seen from the trends.

\begin{figure}
\centering
\includegraphics[width=1\columnwidth]{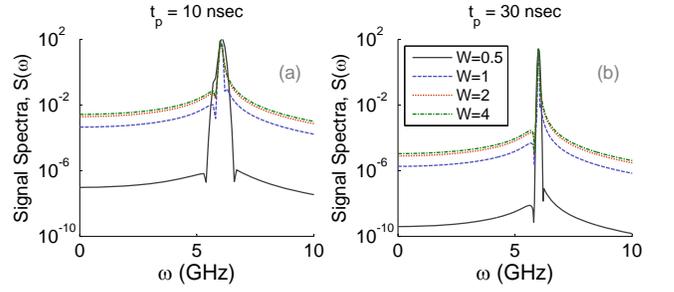}
\caption{ Fourier spectra of DRAG pulses with $\omega=6$ GHz and $A_y=1$ in the lab-frame for pulsewidths of (a) $tp=$ 10 nsec and (b) $t_p=$ 30 nsec for various $W$ cutoffs. }
\label{fig:fft2}
\end{figure}

\section{Results and Discussion}\label{sec.results}

The gate fidelites in this section and in the next are calculated using the fidelity equations in the Appendix (see section.\ref{sec:appendix}). Only two state error fidelites($F^\prime$) are considered in this section.

\subsection{Three-Level System}\label{sec.3L}

\begin{figure}
\centering
\includegraphics[width=1\columnwidth]{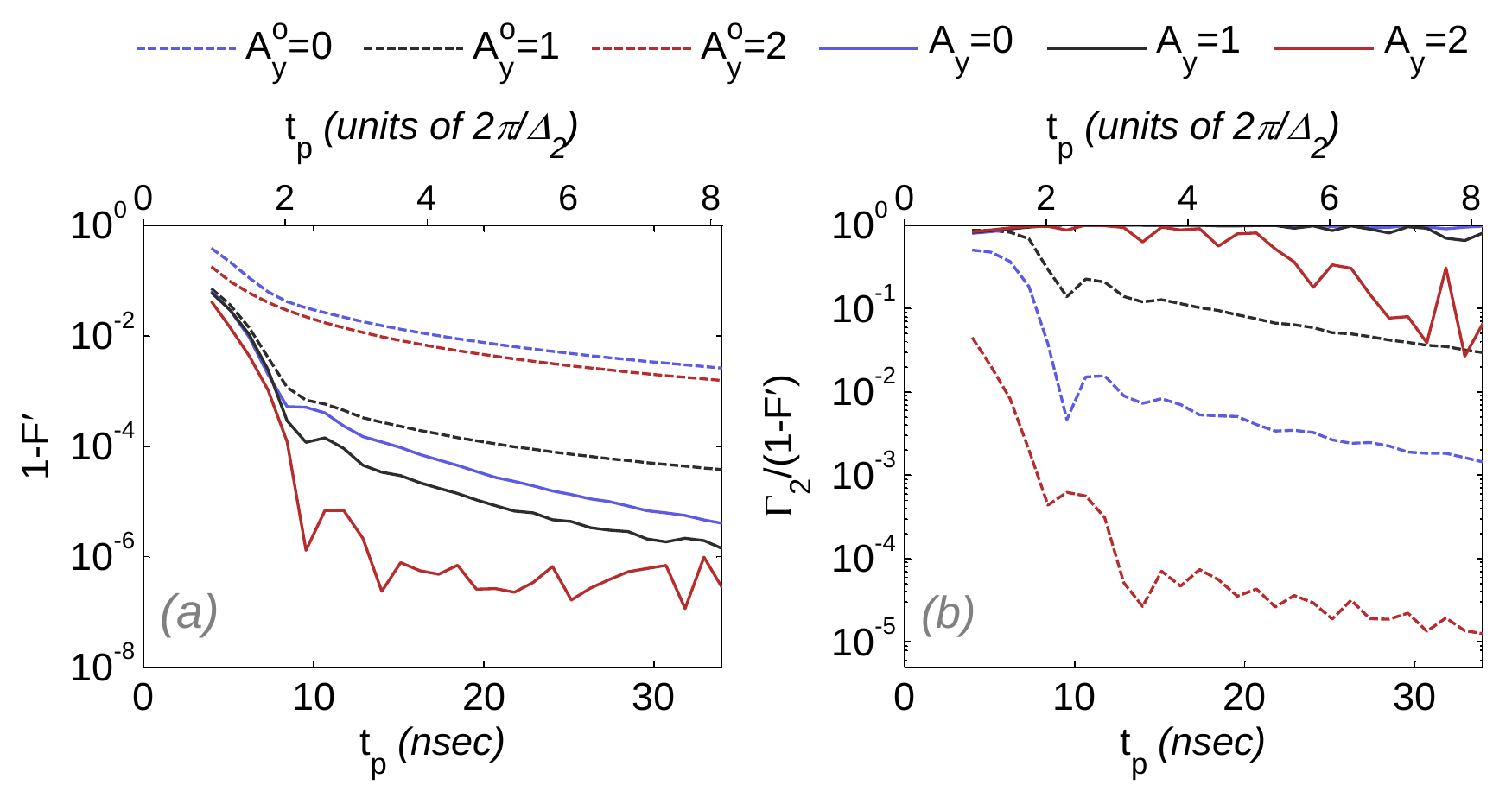}
\caption{{\bf(a)} Two state error infidelity for no-DRAG, $A_y=0$, half-DRAG, $A_y=1$ and full-DRAG $A_y=2$, without any optimization (represented by $A_y^o$) and without optimization (represented by $A_y$) of $A_x$ and $\omega_{01}$. {\bf(b)} Ratio of level-2's population, $\Gamma_2$, to the infidelity ($1-F^\prime$). For all calculations Gaussian pulses with $W=1$ were considered. }
\label{fig:GW1}
\end{figure}

Consider a truncated three-level model.
In fig.\ref{fig:GW1}, the two state error infidelity($1-F^\prime$) for no-DRAG, $A_y=0$, half-DRAG, $A_y=1$ and full-DRAG $A_y=2$ is shown for the cases with numerical optimization and without any optimization (represented by $A_y^o$). For the numerical optimization procedure the drive frequency $\omega_{01}$ and the fundamental pulse amplitude $A_x$ were optimized. The frequency detunning and the pulse amplitude were optimized by iterating over the two parameters serially. The iterations using this taxi-cab method were carried out till a numerical convergence of $10^-8$ was reached. These particular calculations are done for a Gaussian pulse with $W=1$.

It is seen in fig.\ref{fig:GW1}-(b), that when the detuning and the fundamental pulse amplitude are not optimized then the half-DRAG results (represented by the black dashed lines) gives the best improvement in fidelity. This can be understood from a simple spectral analysis where half-DRAG burns the deepest spectral hole. This results in approximately a hundred fold reduction in the infidelity.

However even more dramatic improvements are seen when $\omega$ and $A_x$ are optimized (see Fig.\ref{fig:gausPar} (e) and (f) for the $W=1$ parameters). It is seen here that full-DRAG gives the best results where the infidelity drops to $10^-6$ for $t_p=10$ nsec pulse. This is because the oscillating drive gives rise to AC Stark shift which can be accounted for by detuning $\omega$.  The AC Stark shift energy $dE_s\propto\sqrt{\delta^2-\Omega^2_r}-\delta^2$ where $\delta$ is the detuning and $\Omega_r$ is the Rabi frequency. For longer pulsewidths, $\Omega_r$ is smaller and hence the Stark shift is smaller and hence a smaller detuning is need to account for this. The shift in the energy levels as a function of time pretty much follows the signal $|S(t)|^2$.

Now the shift in the momentum matrix element ,$\bra{0}\Sigma_x\ket{1}$  is inversely proportional to the shift in the energy levels and also follows $|S(t)|^2$. In the absence of DRAG, the shifted $\omega_{\l n}$ levels that are smaller and the corresponding momentum matrix elements are larger due to the shifted wavefunctions. So in the presence of the Stark shift, $A_x$ also needs to be increased in order to maintain the norm of the $\pi$ pulse (see Eq.\ref{eq:Sx}). This is exactly what is seen in the numerically optimized parameters. Once the $A_y$ is increased and the DRAG signal is mixed in, $A_x$ should be decreased and the detuning, $\delta$ follows the same trend.

The average leakage into level-2 at the end of the pulse was also calculated. Qualitatively, $\Gamma_2$ exactly follows the infidelity, $1-F^\prime$. For a quantitative assessment, it is more instructive to examine the ratio of $\Gamma_2$ to $1-F^\prime$ as shown in Fig.\ref{fig:GW1}-(b). This is shown with- and without numerical optimization for Gaussian pulses with $W=1$. The main source of direct leakage is $\mathcal{P}_{1\rightarrow2}$ and also virtual transitions $\mathcal{P}_{0\rightarrow2}$  since $\bra{0}\Sigma_x\ket{2}=0$. The AC Stark shift induced virtual transitions are suppressed by optimizing $\delta$ and $A_x$.

\begin{figure}
\centering
\includegraphics[width=0.9\columnwidth]{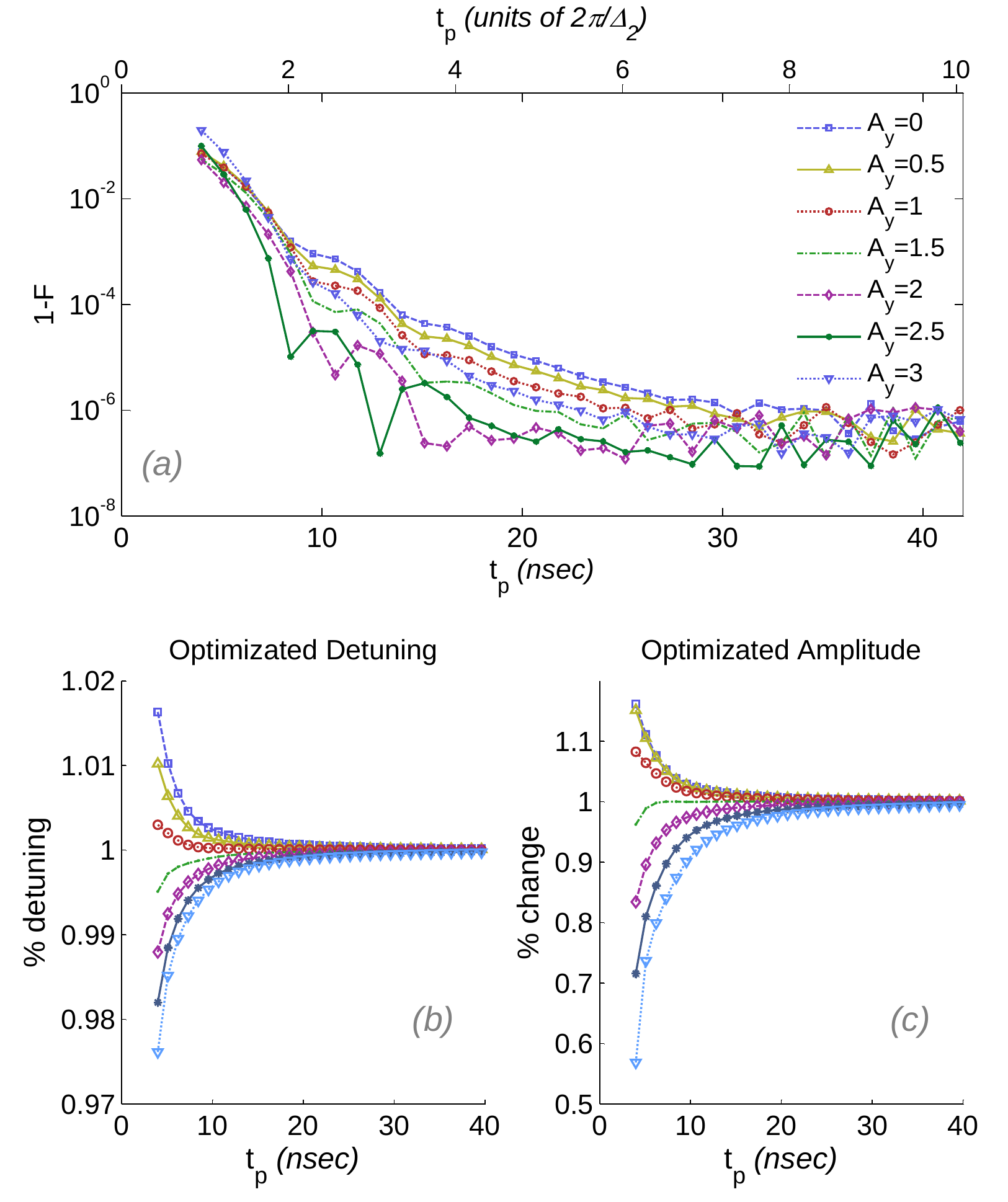}
\caption{(a) Two state error infidelity ($1-F\prime$) for a cosine pulse for different DRAG amplitudes, $A_y$.  The corresponding optimized (b) detuning and (c) fundamental pulse amplitude, $A_x$ is shown as a percentage change. }
\label{fig:cos}
\end{figure}
In order to obtain a better quantitative understanding of the results and to ensure that the parameter space being explored is relevant, it is important to compare the two-state-error fidelities to that obtained by other authors\cite{Martinis2014}. The infidelities for different $A_y$s and the corresponding optimized detuning and fundamental pulse amplitude are shown in Fig.\ref{fig:cos} for the usual $1-\cos$, $\pi$ pulse with $\theta=\pi$, $\alpha_0=-1$ and $\alpha_1=1/2$ in Eq.\ref{eq:cos}. These results  were compared to that shown in Fig.6 of Ref.[\onlinecite{Martinis2014}] and were found to be in excellent agreement for the corresponding $A_y$s (same as $-2D$ in Ref.[\onlinecite{Martinis2014}]). Overall the differences (in the fidelities and optimized parameters) between the $W=1$ Gaussian pulse and the cosine pulse are very small. Subsequent discussions will only focus on Gaussian pulses.

\begin{figure}
\centering
\includegraphics[width=0.9\columnwidth]{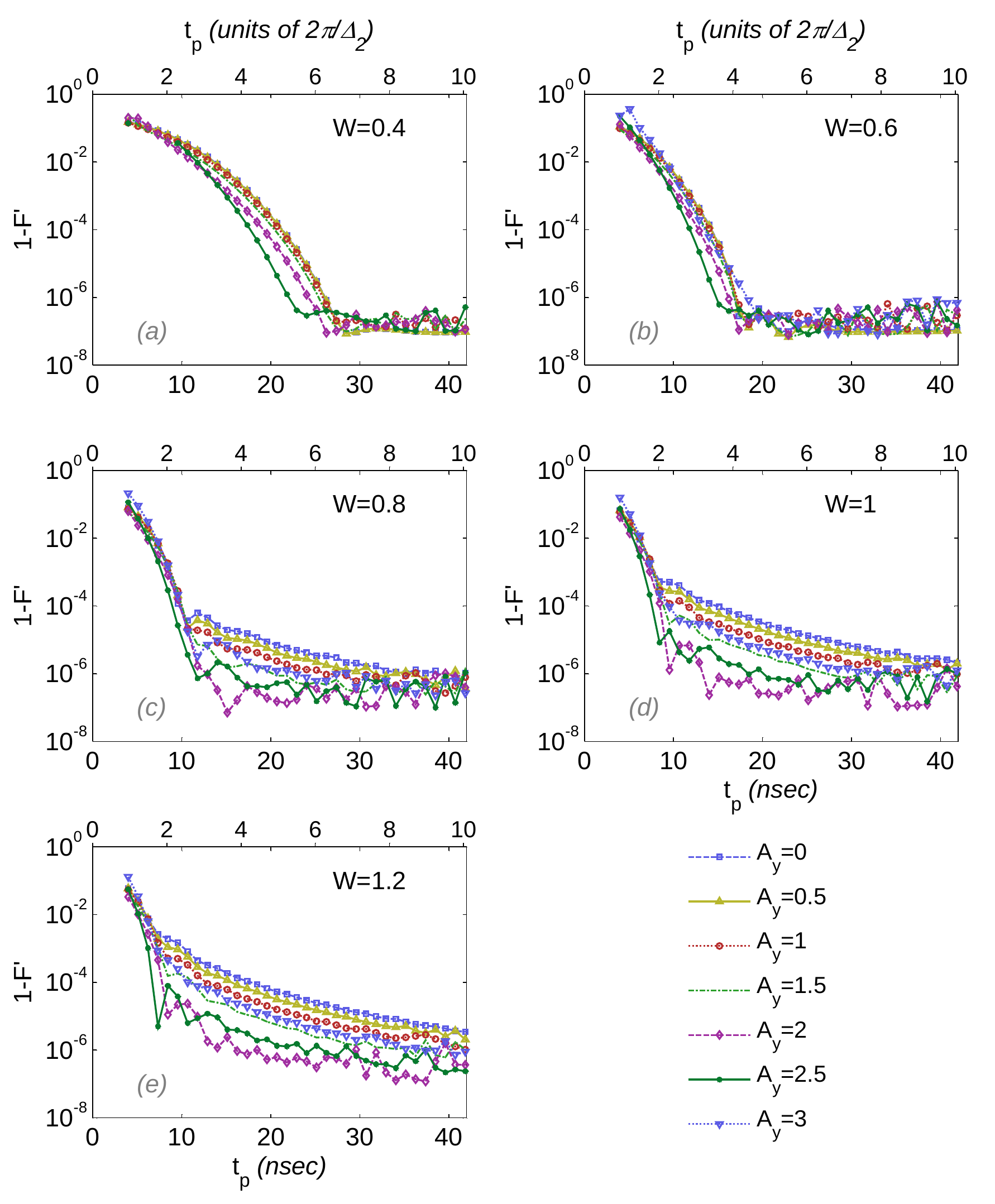}
\caption{ Two state error infidelity ($1-F^\prime$) for different $A_y$s shown for {\bf(a)} $W=0.4$, {\bf(b)} $W=0.6$, {\bf(c)} $W=0.8$, {\bf(d)} $W=1$ and {\bf(e)} $W=1.2$. The frequency detuning and the fundamental pulse amplitude, $A_x$, were optimized. These parameters are shown in fig.\ref{fig:gausPar} for three different cases. }
\label{fig:gausW}
\end{figure}

\begin{figure}
\centering
\includegraphics[width=0.9\columnwidth]{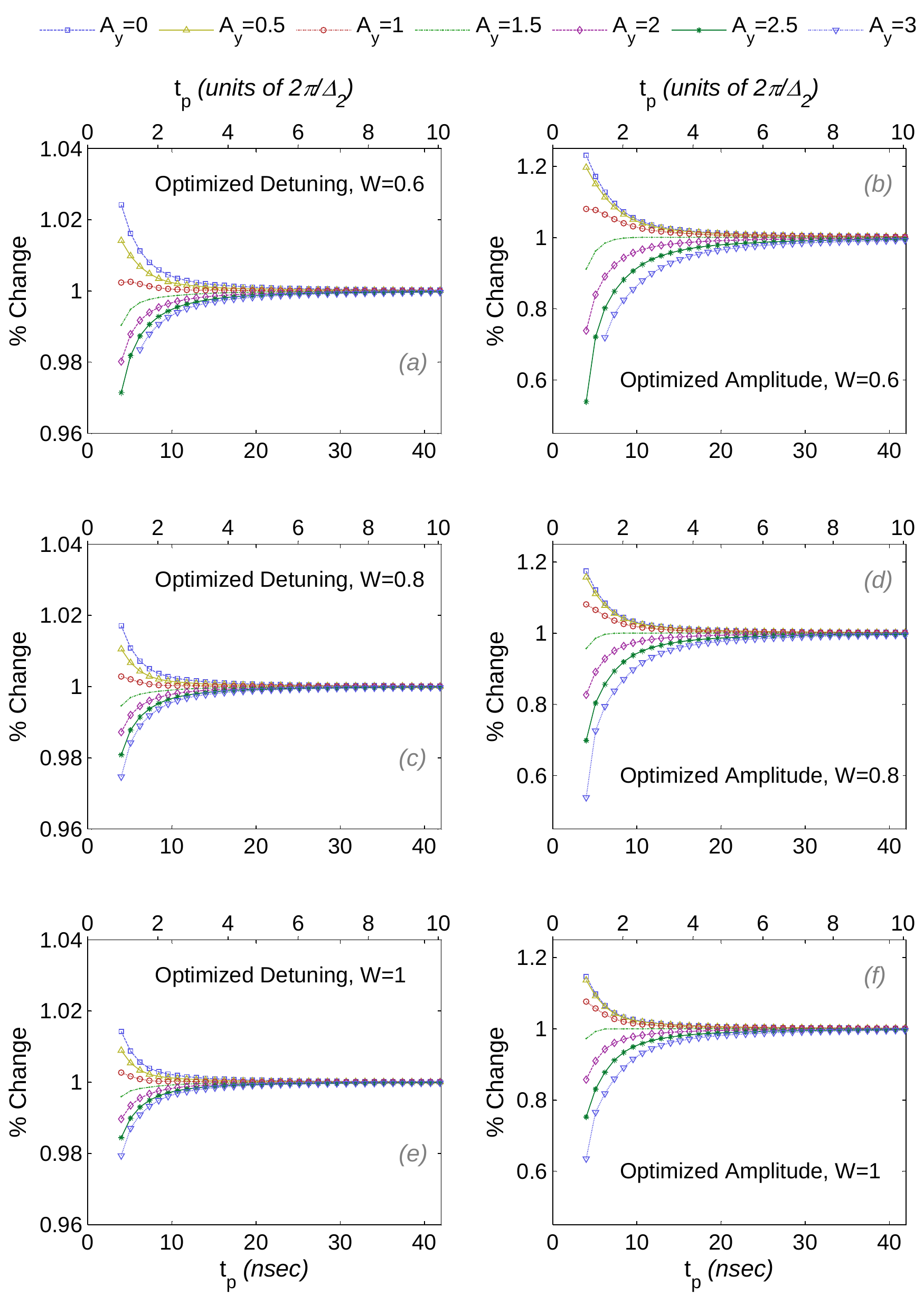}
\caption{Fig.\ref{fig:gausW}'s corresponding optimized detuning and fundamental pulse amplitude, $A_x$ is shown for three different pulse cutoffs $W$. }
\label{fig:gausPar}
\end{figure}

The next part of this discussion focuses on arguably the most important parameter, $W$, which determines the adiabaticity of the pulse. The role of this cutoff parameter has not been discussed in much detail before. And also on as to what happens when the detuning and the pulse amplitude is optimized with- and without DRAG for a more slow rising envelope. Two state error infidelity for different $A_y$s are shown in fig.\ref{fig:gausW} for $W=0.4$-$1.2$. The corresponding optimized frequency detuning and pulse amplitude are shown in fig.\ref{fig:gausPar} for three different cases.

The most striking feature here is that for $W=0.6$ the improvements obtained with DRAG are quite minimal. This is because the spectral-hole burning effect of DRAG plays a minimal role when the spectral weights are low to begin with. Above 20 nsec where the infidelity is flat for $W=0.6$, there appears to be no need for DRAG. Below that limit DRAG can still shorten the pulse times by about $4-5$ nsec. However in this regime the gate fidelities will be very susceptible to fluctuations in $t_p$ due to the sharp slope of $F^\prime$.

For $W=0.4$, the behavior is qualitatively similar, although the pulse times now are longer for stable operations. This is because for $W=0.4$ the resonance linewidth is broader. DRAG with $A_y=2.5$ somewhat improves the gate times by $\sim5$ nsec, but $W=0.4$ pulse is not desirable as over all gates times are much slower. For $W>0.6$, optimized DRAG gives significant improvements with the most fidelity gain occurring for $A_y=2-2.5$. Gate errors of about $\sim 10^-6$ are possible for pulses with $t_p=10$ nsec. This is the fastest gate time. However in this regime the gates would be somewhat susceptible to errors in $t_p$.

\begin{figure}
\centering
\includegraphics[width=0.9\columnwidth]{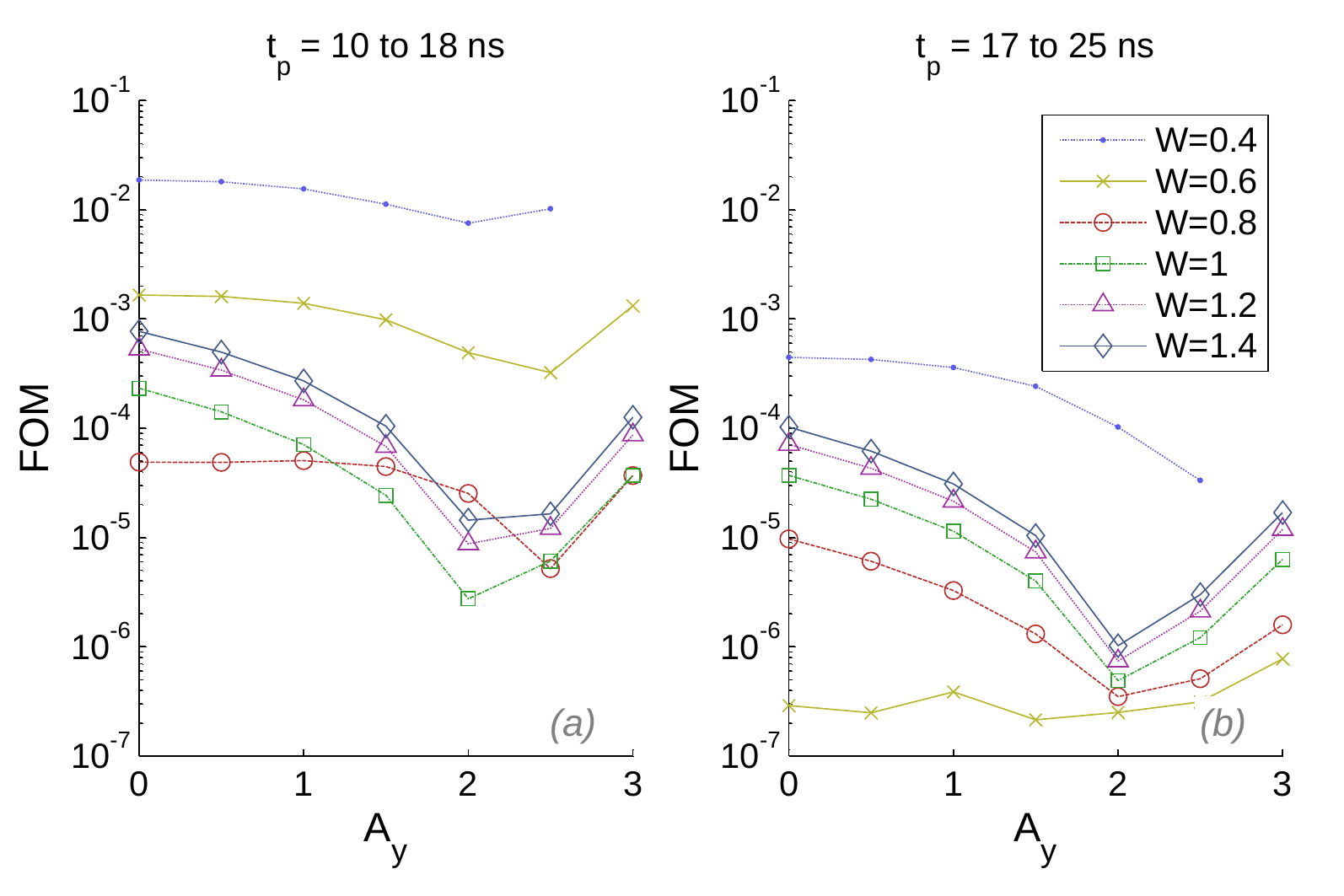}
\caption{ Figure of merit for the Gaussian pulses with various cutoff for $t_p$s between (a)10 to 18 nsec and (b) 17 to 25 nsec.}
\label{fig:FOM}
\end{figure}

%
%

As $F^\prime$ and $dF^\prime/dt_p$ (characterizing the stability) depends on the region of $t_p$, it would be more instructive to characterize the pulses in terms of a figure of merit(FOM) defined as:
\begin{equation}
FOM = 1-\int\limits_{t_{p1}}^{t_{p2}}F^\prime(t_p)dt_p
\end{equation}
The FOM for the Gaussian pulses is shown as a function of $A_y$ for various cutoff and for pulse times between 10 to 18 nsec and 17 to 25 nsec in fig.\ref{fig:FOM}.
For 10 to 18 nsec, $W=1$ and $A_y=2.5$ gives the best fidelities, but this regime is sensitive to pulse errors. More stable operations are possible in the 17 to 25 nsec range with $W=0.6$ where DRAG might not be required.

\subsection{Multi-Level System}\label{sec.8L}
Unfortunately one of the consequences of using a small $W$ Gaussian pulse is that it leads to broader resonance linewidths. In a weakly nonlinear system, a broader linewidth would make the system resonant with other transitions. Irrespective of the initial eigenstate, when the system is driven, the initial state will quickly evolve into a superposition of the states that it is in near resonance with. Hence it is important to go beyond the three-level model to see if there would be any significant changes in the gate fidelities.

\begin{figure}
\centering
\includegraphics[width=1\columnwidth]{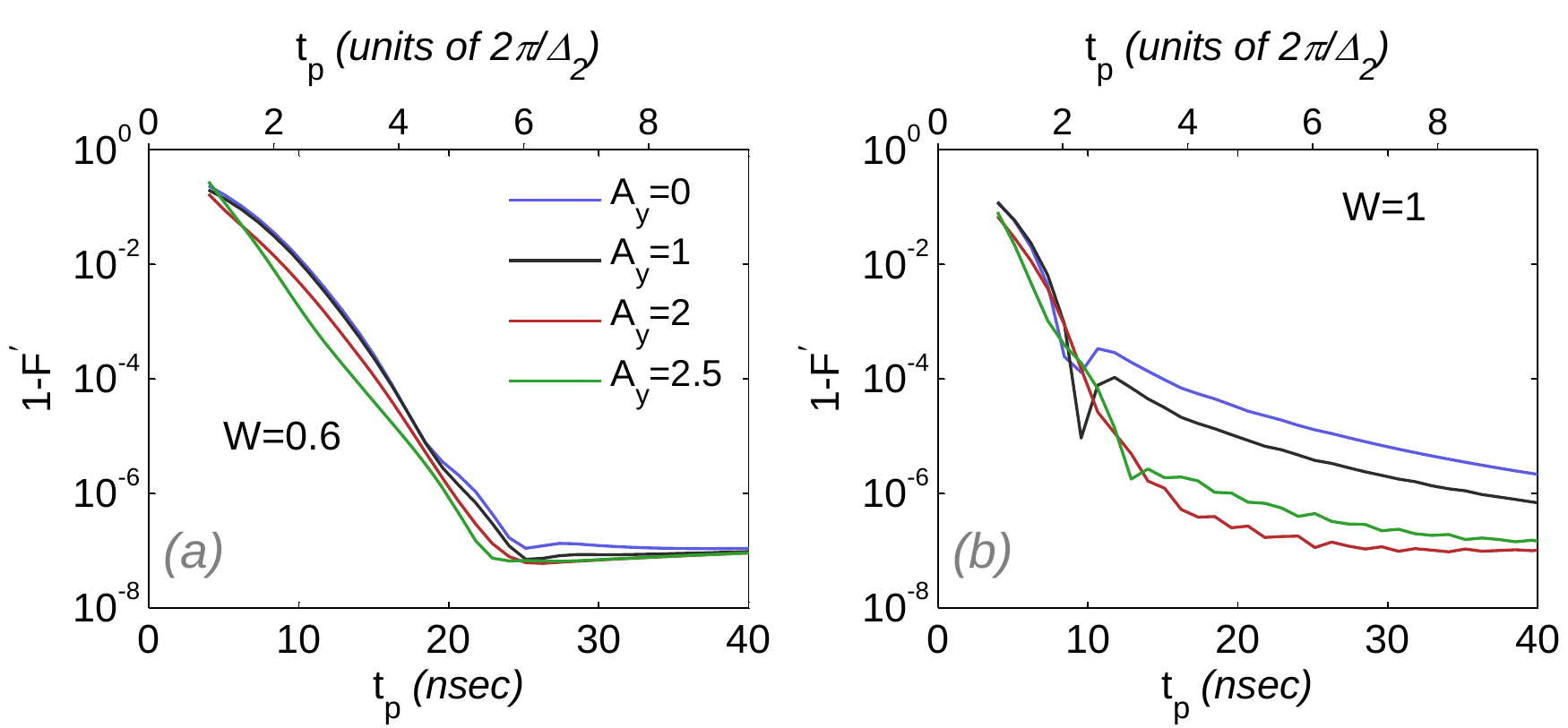}
\caption{ Infidelity, ($1-F^\prime$) for two state errors for 8-level system driven by a Gaussian $\pi$ pulse for different $A_y$ with pulse cut-offs of $W=0.6$ and $W=1$. An 8-level system and a 4-level system have almost identical results.}
\label{fig:8L}
\end{figure}

The calculations were redone for a four-level system and an eight-level system by systematically truncating the tight binding model as described in sec.\ref{sec.model}. There was no difference between the eight-level fidelities and the four-level fidelitites -- the results are identical. The four-level calculations were also repeated with and without the small $\lambda_{03}$ matrix element in $\Sigma'_x$ and there was no difference as the states are far apart in energy. However there are some significant differences between a four-level system and a three-level system which can be attributed to the resonance linewidth.

The two state error infidelity is shown in fig.\ref{fig:8L} for an eight-level system driven by a Gaussian $\pi$ pulse for different $A_y$ with pulse cut-offs of $W=0.6$ and $W=1$. When compared to fig.\ref{fig:gausW}-(b) it is apparent that the addition of the fourth level places greater limitations on the gate times. The shoulders of the $1-F^\prime$ curves are shifted up by roughly $\sim5$ nsec. This is because of additional leakages into the fourth level resulting from the broader resonance linewidth. For $W=1$, the results are very similar to the three-level system results (see fig.\ref{fig:gausW}-(d)) because of the narrower linewidth. Overall, the $W=0.6$ pulses slightly above 20 nsec would be much more stable against pulse jitter. This is a very reasonable gating time, given that the coherence times for Transmons have already hit the sub-ms range\cite{Weides2011,Rigetti2012}.

\section{Summary}\label{sec.summary}

Simultaneously optimizing the detuning and the $\pi$ pulse's norm, to account for the AC Stark shift and the momentum matrix elements respectively, can lead to significantly faster gate times by minimizing the leakage. For Gaussian pulses that have sharper cutoffs (larger $W$), optimized DRAG further improves the fidelity. Gate infidelties of $\sim10^{-5}$ are possible for 10 nsec pulses, although this is not a very stable regime. Much more stable gate operations are possible at $t_p\sim20$ nsec with smaller $W$ pulses with lower overall spectral weights but where $t_p^{min}$ is more limited by the resonance linewidth. For a more adiabatic envelope such as this, optimized DRAG makes only a small difference.

Although for smaller spectral weight pulses with fewer Fourier components are best for reducing leakages into unwanted levels for an anharmonic oscillator, for certain applications such as dynamical-decoupling(DD), pulses with more fourier components do exceedingly when it comes to cancelling unwanted interactions and noise. For example, it was shown\cite{De2014pra} that in the presence of Ising interactions and dephasing noise for ideal spin-1/2 qubits, second order self-refocusing pulses can improve one- and two-qubit gate fidelities by an additional four orders of magnitude when compared to Gaussian pulses.

In order to address the wide range of issues associated with gating a qubit more careful pulse design considerations are needed. Ideally the pulses should be designed to minimize two state errors, minimize phase errors, minimize noise and be able to cancel unwanted qubit-qubit interactions. This would however be a challenging task and would require requires careful and detailed system models.

\section{Acknowledgements}

I wish to thank Alexander Korotkov for his initial participation and for a number of key discussions. I would also like to thank Leonid Pryadko for several very helpful discussions and for his support. AD has been supported by the U.S. Army Research Office, Grant No.~W911NF-11-1-0027 and  W911NF-14-1-0272, and by the NSF, Grant No.~1018935.

\section{Appendix: Calculating Fidelities}\label{sec:appendix}

More details are given here for the time-ordered time evolution calculations.  A simple prescription is given for numerically calculating fidelities for all $x,y$ and $z$ terms in the lab-frame for a multi-level model. As the phase terms oscillate with the drive, one has to go into the counter rotating frame in a time-ordered manner.

\begin{figure}
\centering
\includegraphics[width=1\columnwidth]{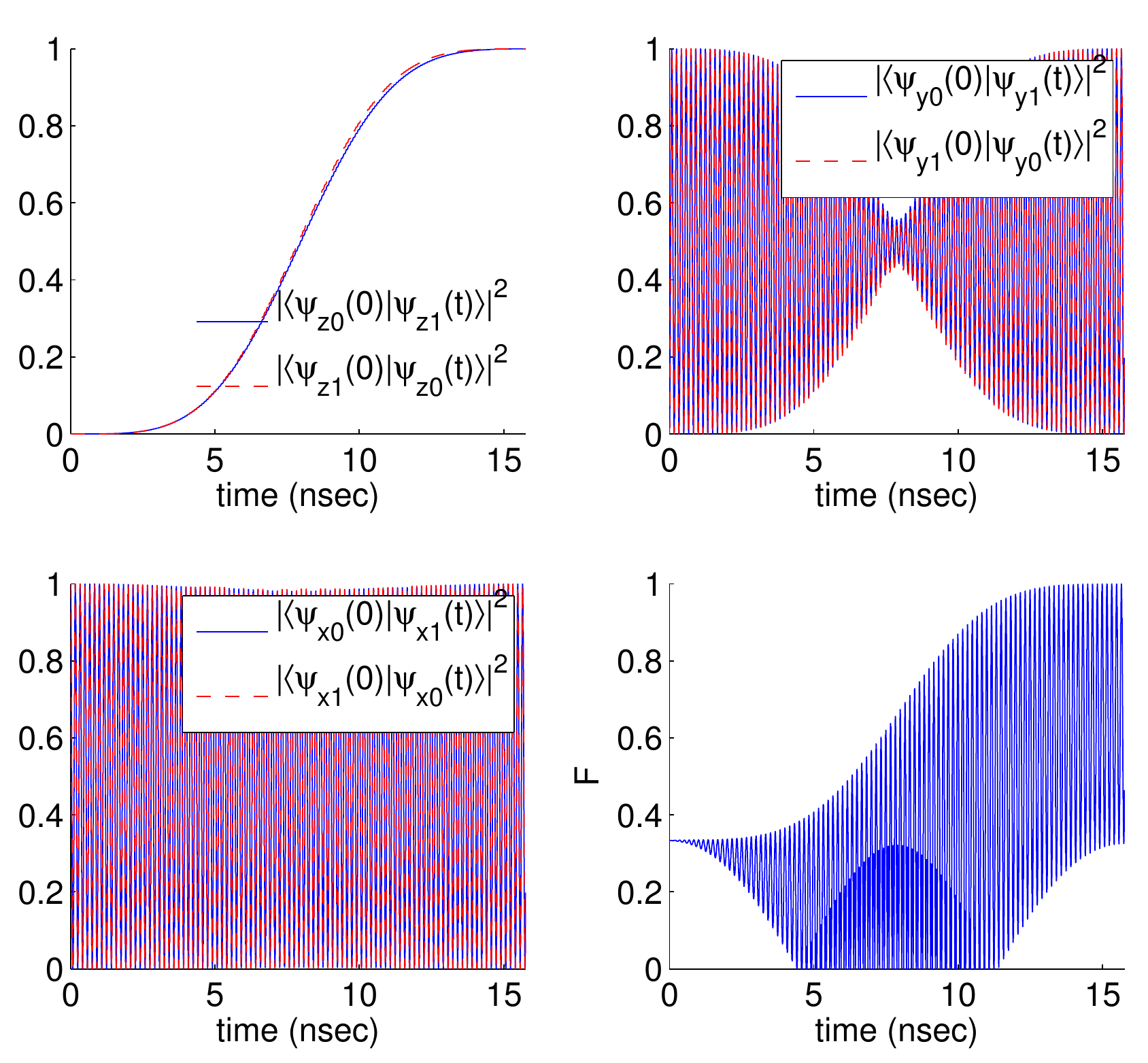}
\caption{ Transitions in the lab frame between $\ket{0}_z\leftrightarrow\ket{1}_z$, $\ket{-}_y\leftrightarrow\ket{+}_y$ and $\ket{-}_x\leftrightarrow\ket{+}_x$ and the full fidelity for a Gaussian DRAG $\pi$ pulse applied along $x$.}
\label{fig:lframe}
\end{figure}

\begin{figure}
\centering
\includegraphics[width=1\columnwidth]{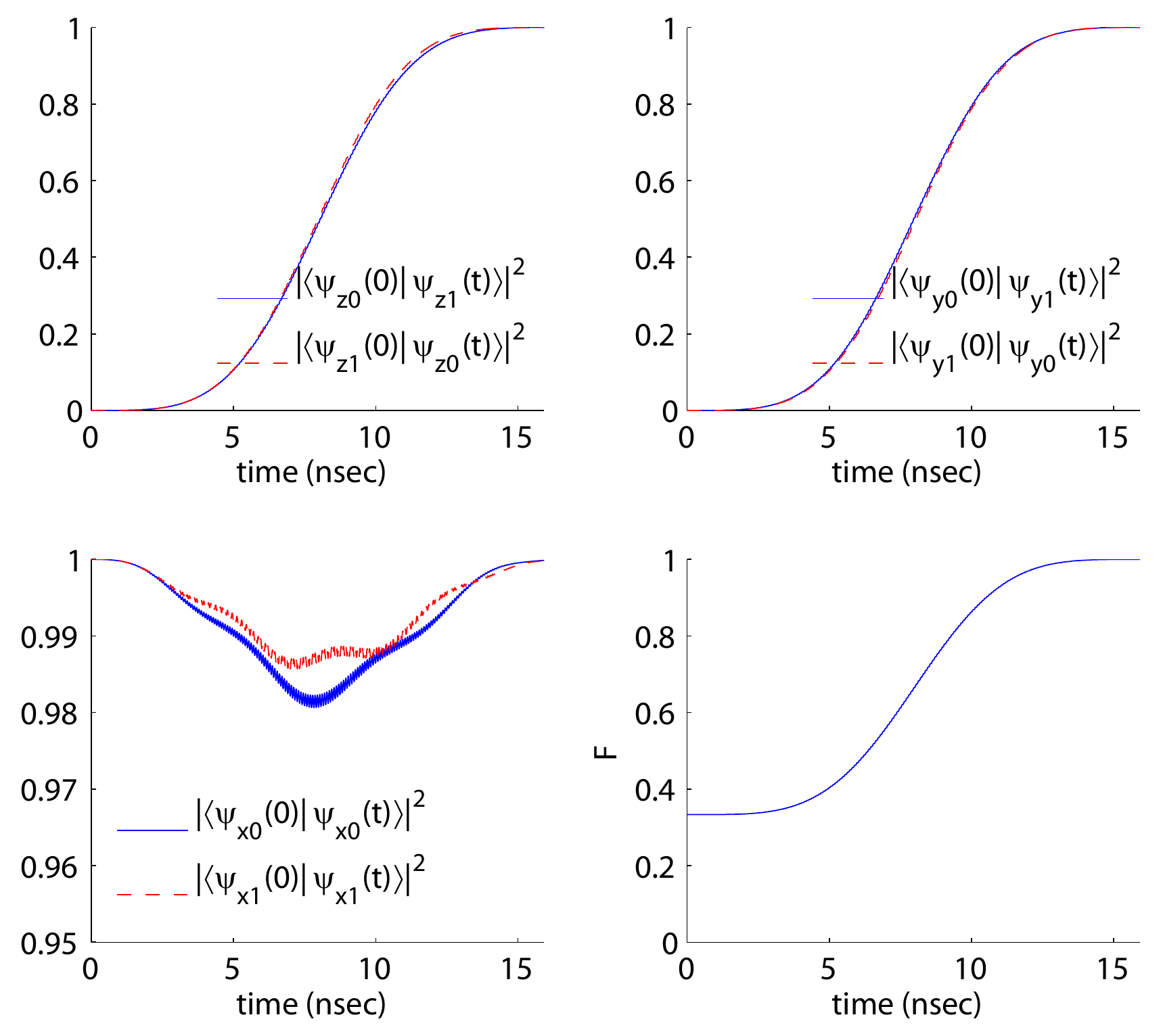}
\caption{ Transitions in the rotating frame between $\ket{0}_z\leftrightarrow\ket{1}_z$, $\ket{-}_y\leftrightarrow\ket{+}_y$ and $\ket{-}_x\leftrightarrow\ket{+}_x$ and the full fidelity for a Gaussian DRAG $\pi$ pulse applied along $x$.}
\label{fig:rframe}
\end{figure}

Consider a truncated 3-level model. The eigenvectors are:
\begin{eqnarray}
\ket{0}_z = [1,0,0]\\
\ket{1}_z = [0,1,0]\\
\ket{2}_z = [0,0,1]
\label{eq:z}
\end{eqnarray}
The eigenvectors for $x$ and $y$ can be constructed as follows:
\begin{eqnarray}
\ket{\pm}_x &=& \left(\ket{0}_z\pm\ket{1}_z\right)/\sqrt{2}\\
\ket{\pm}_y &=& \left(\ket{0}_z\mp i\ket{1}_z\right)/\sqrt{2}
\label{eq:z}
\end{eqnarray}
If one starts with $\ket{\Psi_o}=\{ \ket{+}_x,\ket{-}_x,\ket{+}_y,\ket{-}_y,\ket{0}_z,\ket{1}_z \}$, then the full fidelity at any time $t$ is:
\begin{eqnarray}
F&=&\frac{1}{{N}}Tr\left[|\mathcal{M}|^2\right]\\
\mathcal{M}&=&\langle\Psi_e|U(t_n)...U(t_1)U(t_0)|\Psi_o\rangle
\label{eq:F}
\end{eqnarray}
where the time evolution is carried out in $n$ discretized time steps and $U(t_k)=\exp(-\frac{i}{\hbar}H(t_k)dt)$. Here $\ket{\Psi_e}$ comprises of column vectors of the expected wavefunctions (depending on the operation) and ${N}$ is the number of column vectors (for the full fidelity $\mathcal{N}=6$).

However when the fidelity is calculated this way in the lab-frame, it leads to rapid oscillations for $x$ and $y$ components as seen in fig.\ref{fig:lframe}. An easy fix for this is to go into the counter rotating frame as follows:
\begin{eqnarray}
\mathcal{M}=\langle\Psi_e|e^{iH_ot_n}U(t_n)...e^{iH_ot_1}U(t_1)e^{iH_ot_0}U(t_0)|\Psi_o\rangle~~
\label{eq:F2}
\end{eqnarray}
Note that that the choice between $e^{+iH_ot}$ or $e^{-iH_ot}$ to go into the counter rotating frame depends on the sign of the argument of $U$. The one with the opposite sign fixes the problem with the rapidly oscillating phase terms as shown in fig.\ref{fig:rframe}

For the \emph{two-state-error} fidelity: $\ket{\Psi_o}=\{\ket{0}_z,\ket{1}_z \}$ and $\mathcal{N}=2$. All calculations in this paper are for the two-state-error fidelity denoted by $F^\prime$.



\end{document}